\documentclass[aps,prd,notitlepage,showpacs,nofootinbib,superscriptaddress,12pt]{revtex4-1}

\usepackage{graphicx} % Required for inserting images

\usepackage[T1]{fontenc}
\usepackage{lmodern}
\usepackage{caption}
\usepackage{floatrow}
\usepackage{subcaption}
\usepackage{hyperref}
\hypersetup{colorlinks,linkcolor={blue},citecolor={blue},urlcolor={blue}}
\usepackage[normalem]{ulem}

\usepackage{color}

\usepackage{mathtools}
\usepackage{physics}
\usepackage{amsmath}
\usepackage{amssymb}
\usepackage{slashed}

\begin{document}

\title{Constraining the isotropic $CPT-$odd coefficients of the Standard Model Extension by a combined DUNE and ESSnuSB analysis}

\author{Luis A. Delgadillo} \email{ldelgadillof2100@alumno.ipn.mx}
\affiliation{Departamento de F\'{\i}sica, Escuela Superior de
  F\'{\i}sica y Matem\'aticas del Instituto Polit\'ecnico Nacional, Unidad Adolfo L\'opez Mateos, Edificio 9, 07738 Ciudad de M\'exico, Mexico}
\affiliation{Institute of High Energy Physics, Chinese Academy of Sciences, Beijing 100049, China}  
\author{O.~G.~Miranda} \email{omar.miranda@cinvestav.mx}
\affiliation{Departamento de F\'{\i}sica, Centro de Investigaci{\'o}n y de Estudios Avanzados del IPN
  Apdo. Postal
  14-740 07000 Ciudad de M\'exico, Mexico}
\author{G.~Moreno-Granados} \email{guadalupe.moreno@cinvestav.mx}
\affiliation{Departamento de F\'{\i}sica, Centro de Investigaci{\'o}n y de Estudios Avanzados del IPN
  Apdo. Postal
  14-740 07000 Ciudad de M\'exico, Mexico}
\affiliation{Center for Neutrino Physics, Virginia Tech, Blacksburg, VA, 24061, USA}
\author{C.~A.~Moura}%
\email{celio.moura@ufabc.edu.br}
\affiliation{Centro de Ci\^encias Naturais e Humanas, Universidade Federal do ABC - UFABC, Av. dos Estados, 5001, 09210-580, Santo Andr\'e-SP, Brazil}

\date{\today}

\begin{abstract}
\noindent Based on an analysis that considers the isotropic $CPT-$odd Standard Model Extension (SME) coefficients, we find new constraints for them coming from a combined DUNE and ESSnuSB fit. Furthermore, we investigate the correlations of the standard oscillation parameters, the leptonic $CP-$violating phase, $\delta_{CP}$, and the atmospheric mixing angle, $\sin^2 \theta_{23}$, with respect to the SME coefficients $(a_L)^T$. The combination of DUNE and ESSnuSB may establish the strongest limit on the diagonal coefficient $(a_L)_{\mu \mu}^T$ compared to the existing limits in the literature. We also consider the possible effect of the SME coefficient $(a_L)^Z$ on neutrino propagation and discuss how this can affect DUNE limits on the coefficient $(a_L)^T$ found elsewhere.
\end{abstract}

\maketitle

\section{Introduction}
\label{intro}
On the neutrino oscillation frontier, where nearly all of the available data is compatible with the standard three-flavor oscillation paradigm, precision measurements can be performed to investigate Beyond Standard Model (BSM) physics that may disturb the oscillation pattern. Looking for perturbations on the neutrino oscillation standard model brings new opportunities to access, for example, the Planck scale.

In the three neutrino oscillation picture, NOvA~\cite{NOvA:2021nfi, NOvA:2023iam}, T2K~\cite{T2K:2023smv, T2K:2023mcm}, and recently IceCube~\cite{IceCube:2024xjj} collaborations are leading the precision measurements on both the atmospheric oscillation parameters, $\Delta m^2_{32}$ and $\sin^2 \theta_{23}$, and the $CP-$violating phase, $\delta_{CP}$~\cite{T2K:2023smv}.
The current measurement of the angle $\theta_{23}$ demonstrates consistency with maximal mixing (that is, $\theta_{23}=\pi/4$). 

This has implications for our understanding of neutrino masses and the nature of neutrino flavor mixing.
Furthermore, the precise determination of $\sin^2 \theta_{23}$ is crucial for refining our understanding of the neutrino mass hierarchy, the order of neutrino mass eigenstates, and for probing possible sources of $CP$ violation in the neutrino sector.

On the other hand, accurately determining $\delta_{CP}$ is a primary goal of several ongoing and planned long-baseline neutrino oscillation experiments. Present$-$day measurements of $\delta_{CP}$ have been reported, mainly by NOvA and T2K collaborations. Their results seem to differ and might be a challenge to interpret if this difference persists in the future. Although the best fit of the NOvA collaboration agrees with a value of $\delta_{CP} \simeq 0.8 \pi$~\cite{NOvA:2021nfi}, a similar analysis of the T2K collaboration reports $\delta_{CP} \simeq 1.4 \pi$~\cite{T2K:2023smv}, under Normal Order (NO) of the neutrino mass. 
A recent joint oscillation analysis of the Super-Kamiokande and T2K collaborations is consistent with a best-fit value of $\delta_{CP} \simeq 1.4 \pi$~\cite{T2K:2024wfn}.
However, some proposals to alleviate this discrepancy in the measurement of $\delta_{CP}$, which has a statistical significance of $\sim 2\sigma$ CL~\cite{Chatterjee:2024kbn}, include nonstandard neutrino interactions (NSI)~\cite{Denton:2020uda, Chatterjee:2020kkm, Delgadillo:2023lyp}, sterile neutrinos~\cite{Chatterjee:2020yak, deGouvea:2022kma}, among other BSM scenarios~\cite{Lin:2023xyk}.

The Standard Model of Particle Physics (SM) is thought to represent a low-energy effective gauge theory of a more fundamental theory in which the Planck scale ($M_{\text{Pl}} \sim 10^{28}$ eV) would be the natural mass scale. Likewise, the Standard Model Extension (SME) framework~\cite{Colladay:1998fq} parameterizes the possible breakdowns of the Lorentz and $CPT$ symmetries that might arise from a fundamental theory that incorporates both the gravitational force from one side and the strong, weak, and electromagnetic interactions of the SM from the other side. 
Such potential violations of the $CPT$ and Lorentz symmetries may arise at very high energies, well above the electroweak scale, for instance, within the context of string theory~\cite{Kostelecky:1988zi, Kostelecky:1991ak, Kostelecky:1995qk, Colladay:1996iz}.
Furthermore, it has been argued that a potential Lorentz invariance violation (LIV) and symmetry breakdown $CPT$ in the neutrino sector might result from NSI of neutrinos with scalar fields~\cite{Gu:2005eq, Ando:2009ts, Simpson:2016gph, Klop:2017dim, Capozzi:2018bps, Farzan:2018pnk, Gherghetta:2023myo, Arguelles:2023wvf, Arguelles:2023jkh, Lambiase:2023hpq, Cordero:2023hua, Arguelles:2024cjj}.

For reviews on $CPT$ violation and LIV in the neutrino sector, see, e.g., Refs.~\cite{Kostelecky:2003cr, Kostelecky:2011gq, Diaz:2016xpw, Torri:2020dec, Moura:2022dev, Barenboim:2022rqu}.
As far as neutrino oscillation experiments are concerned, $CPT$ violation in the neutrino sector was suggested to explain the Liquid Scintillator Neutrino Detector (LSND) and  MiniBooNE anomalies \cite{Murayama:2000hm, Barenboim:2001ac, Barenboim:2004wu, Kostelecky:2004hg, MiniBooNE:2011pix, Katori:2012pe}.
For studies on possible LIV and $CPT$-violating effects in various neutrino oscillation experiments, we refer to~\cite{Bahcall:2002ia, DoubleChooz:2012eiq, MINOS:2008fnv, MINOS:2010kat, IceCube:2010fyu, Super-Kamiokande:2014exs, SNO:2018mge, Barenboim:2017ewj, Barenboim:2018ctx, KumarAgarwalla:2019gdj, Rahaman:2021leu, Ngoc:2022uhg, Sarker:2023mlz, Raikwal:2023lzk, Sahoo:2021dit, Agarwalla:2023wft, Mishra:2023nqf, Barenboim:2018lpo, Sahoo:2022nbu, Majhi:2022fed, Barenboim:2023krl, Pan:2023qln, Arguelles:2015dca, Fiza:2022xfw, Cordero:2024nho, Shukla:2024fnw, Mishra:2024vja, IceCube:2021tdn, Testagrossa:2023ukh, Telalovic:2023tcb}.
Moreover, non-oscillatory probes of Lorentz and $CPT$ breakdowns in the neutrino sector include: beta decay~\cite{Diaz:2013saa, Lehnert:2021tbv, KATRIN:2022qou}, double beta decay~\cite{EXO-200:2016hbz, CUPID:2019kto}, and cosmic neutrino background~\cite{Diaz:2015aua}.

Long-baseline neutrino oscillation experiments are crucial for solving puzzles in the conventional three-neutrino paradigm as well as for investigating different new physics scenarios, such as possible breakdowns of the Lorentz and $CPT$ symmetries. Other works have studied the effects of LIV on the sensitivity of DUNE, as for example Refs.~\cite{Barenboim:2017ewj, Barenboim:2018ctx, KumarAgarwalla:2019gdj}. However, in this analysis, we employ the most recent DUNE technical design report (TDR) configuration files provided by the collaboration~\cite{DUNE:2021cuw}.

In this paper, we examine the potential violations of Lorentz and $CPT$ symmetries within the context of future accelerator-based long-baseline neutrino oscillation experiments. We examine the following in particular: 1. The Deep Underground Neutrino Experiment (DUNE)~\cite{DUNE:2020jqi}, a next-generation long-baseline neutrino oscillation experiment, and 2. The next-to-next generation neutrino oscillation experiment, the European Spallation Source neutrino Super-Beam (ESSnuSB) at the European Spallation Source (ESS)~\cite{Alekou:2022emd}. We show how a combined analysis of these two experiments may improve the sensitivity to Lorentz and $CPT$ studies compared to DUNE alone.

\section{Theoretical framework and experimental configurations}
\label{framework}

We describe the neutrino propagation from the source to the detector considering the Hamiltonian containing three components,
\begin{equation}
   \hat{H}=\hat{H}_{\text{vacuum}} + \hat{H}_{\text{matter}} + \hat{H}_{\text{LIV}}\,,
\end{equation}
where $\hat{H}_{\text{vacuum}}$ describes the neutrino propagation in vacuum, $\hat{H}_{\text{matter}}$ contains the Mikheyev-Smirnov-Wolfenstein (MSW) matter potential, and $\hat{H}_{\text{LIV}}$ includes the perturbative effects from LIV and possibly $CPT$ symmetry violation. Explicitly, considering only the relevant $CPT-$odd coefficients of the SME neutrino sector,
\begin{equation}
\label{LIVHAM}
 \hat{H}_{\text{LIV}} =   \left(
\begin{array}{ccc}
 a_{ee} & a_{e \mu} & a_{e \tau} \\
 a_{e \mu}^{*} & a_{\mu \mu} &  a_{ \mu \tau} \\
 a_{e \tau}^{*} & a_{ \mu \tau}^{*} & a_{ \tau \tau} \\
\end{array}
\right) \,,
\end{equation}
where $a_{\alpha\beta} = (a_L)^T_{\alpha\beta}$ represent the time component of the coefficients for each neutrino oscillation sector ($\alpha,\beta=e,\mu,\tau\,$). We neglect the script $T$ to simplify the notation.

Next, we present the DUNE and ESSnuSB experiments, which will serve as the platforms for studying the potential effects of Lorentz invariance and $CPT$ symmetry violations. However, in Appendix~\ref{angles}, we consider the correlations of time $T$ and $Z-$spatial components of the $CPT-$odd SME coefficients at the DUNE configuration.

\subsection{DUNE}
To simulate DUNE, we used the DUNE configuration ancillary files available~\cite{DUNE:2021cuw} for GLoBES and the specifications of the DUNE (TDR)~\cite{DUNE:2020jqi,DUNE:2020lwj, DUNE:2020ypp}, where more details and information about the experiment can be found. DUNE will consist of up to four modules, each containing approximately 10 kton liquid argon detectors, with a mean neutrino energy around 2.5~GeV, located at 1285 km from the beam source (on axis) with a 1.2 MW power in the first phase of operations. We consider a time exposure of ten years, evenly distributed between the neutrino and antineutrino modes. The detectors will be liquid argon time projection chambers (LArTPC), which can track charged particles created after neutrino interaction. Information on mixing angles, flux, and their uncertainties is given in section~\ref{sec:method}. Neutrinos are considered to travel through a constant matter density equal to 2.848~g/cm$^3$.

\subsection{ESSnuSB}
The ESSnuSB proposal, a long-baseline neutrino experiment next-to-next generation, will consist of roughly 538 kt of water Cherenkov detector located underground of the Zinkgruvan mine in Sweden~\cite{Alekou:2022emd}. In this analysis, we follow the experimental configuration and systematic uncertainties as described in Refs.~\cite{Delgadillo:2023lyp, Cordero:2022fwb, Cordero:2024nho}, which consider a baseline $L=360$ km and a ten-year exposure divided equally between neutrino and antineutrino modes; the matter density was assumed to be 2.8~g/cm$^3$. The neutrino beam (on-axis) intensity corresponds to a 2.0 GeV proton beam with $2.7\times10^{23}$ protons on target per year~\footnote{The expected annual operation period is estimated to be 208 days.} fixed at 5 MW power~\cite{Cordero:2022fwb}. The neutrino fluxes have been properly rescaled to a recent simulation with 2.5 GeV proton kinetic energy~\cite{Alekou:2022emd}. Besides, in this study, we consider an energy resolution which follows a Gaussian distribution, with a width of $\sigma(E)=12\%/\sqrt{E [\mbox{GeV}]}$ for electrons and $\sigma(E)=10\%/\sqrt{E [\mbox{GeV}]}$ for muons, accordingly. A total of 12 bins uniformly distributed in the energy interval of $0.1-1.3$ GeV were considered~\cite{Cordero:2022fwb}. For further details on the experimental setup and physics goals of ESSnuSB, we refer the reader to~\cite{Alekou:2022emd, ESSnuSB:2013dql, Blennow:2019bvl, Capozzi:2023ltl, ESSnuSB:2024wet}.

\section{Methodology}
\label{sec:method}
In this study, we use GLoBES~\cite{Huber:2004ka, Huber:2007ji} and its additional NSI plug-in \emph{snu.c}~\cite{Kopp:2006wp, Kopp:2007rz}, which was modified to implement the $CPT-$odd SME coefficients in the Hamiltonian. 
We employ a minimum-square analysis to quantify the statistical significance of the $CPT-$odd SME coefficients, using both neutrino and antineutrino datasets. The total $\chi^2-$function is given as~\cite{Delgadillo:2023lyp}
\begin{equation}
    \chi^2 = \sum_{\ell} \tilde{\chi}^2_{\ell} + \chi^2_{\text{prior}},
\end{equation}
where the corresponding $\tilde{\chi}^2_{\ell}-$function stands for each channel $\ell$, with $\ell= \big( \nu_{\mu}(\Bar{\nu}_{\mu})\rightarrow \nu_{e} (\Bar{\nu}_{e}),~\nu_{\mu}(\Bar{\nu}_{\mu})\rightarrow \nu_{\mu} (\Bar{\nu}_{\mu}) \big)$, and is provided as in Ref.~\cite{Huber:2002mx}
\begin{equation}
\begin{split}
    &\tilde{\chi}_{\ell}^2= \min_{\xi_{j}} \Bigg[  \sum _{i}^{n_{\text{bin}}} 2 \Bigg\{ N_{i,\text{test}}^{3 \nu+\text{LIV}}( \Omega, \Theta, \{\xi_{j}\})-N_{i,\text{true}}^{3\nu} + N_{i,\text{true}}^{3\nu} \log \frac{N_{i,\text{true}}^{3\nu}}{N_{i,\text{test}}^{3 \nu+\text{LIV}}( \Omega, \Theta, \{\xi_{j}\})} \Bigg\} \\
    &~~~~~~~~~~~~~~~ + \sum_{j}^{n_{\text{syst}}} \Big(\frac{\xi_{j}}{\sigma_{j}}\Big)^2 \Bigg],
\end{split}
\end{equation}
where $N_{i, \text{true}}^{3\nu}$ are the simulated events at the $i$-th energy bin, considering the standard three neutrino oscillations framework, $N_{i, \text{test}}^{3\nu +\text{LIV}}( \Omega, \Theta, \{\xi_{j}\})$ are the computed events at the $i$-th energy bin including $CPT-$odd SME coefficients (one parameter at a time). In addition, $\Omega = \{\theta_{12}, \theta_{13}, \theta_{23}, \delta_{CP}, \Delta m_{21}^2, \Delta m^2_{31}\}$ is the set of neutrino oscillation parameters, while $\Theta = \{|a_{\alpha \beta}|, \phi_{\alpha\beta}, a_{\alpha \alpha}, \cdots \}$ is the set of isotropic SME coefficients, $a_{\alpha \beta}$, and $\{\xi_{j}\}$ are the nuisance parameters to account for systematic uncertainties. Moreover, $\sigma_{j}$ are the systematic uncertainties; for the DUNE setup, as reported in the DUNE TDR~\cite{DUNE:2020ypp}, and for the ESSnuSB configuration, we assume a 10\% signal normalization uncertainty and 15\% background normalization error for both appearance and disappearance channels. In addition, a $0.01\%$ energy calibration uncertainty has been considered as in Ref.~\cite{Cordero:2022fwb}. To obtain our simulated events, we consider the neutrino oscillation parameters of Salas \emph{et al.}~\cite{deSalas:2020pgw} as~\emph{true} values, shown in Table~\ref{tab:osc}.

\begin{table}[ht]
\centering
\caption{\label{tab:osc}Standard oscillation parameters used in our analysis~\cite{deSalas:2020pgw}. We consider the NO throughout this study.}
\begin{tabular}{c  c}
\hline \hline
Oscillation parameter & Best fit NO  \\
\hline 
$\theta_{12}$ & 34.3$^{\circ}$ \\
$\theta_{23}$ & 49.26$^{\circ}$ \\
$\theta_{13}$ &  8.53$^{\circ}$ \\
$\Delta m^2_{21}$ [10$^{-5}$~eV$^2$] & 7.5  \\ 
$|\Delta m_{31}^2|$ [10$^{-3}$~eV$^2$] & 2.55  \\ 
$\delta_{CP}/ \pi$ & 1.08  \\
\hline \hline
\end{tabular}
\end{table} 

Furthermore, the implementation of external input for the standard oscillation parameters on the $\chi^2$ function is performed via Gaussian priors~\cite{Huber:2002mx}
\begin{equation}
    \chi^2_{\text{prior}}= \sum_{k}^{n_{\text{priors}}}   \frac{\big(\Omega_{k,\text{true}}-\Omega_{k,\text{test}}\big)^2}{\sigma^2_{k}}\,. 
\end{equation}
The central values of the oscillation parameter priors, $\Omega_{k,\text{true}}$, are fixed to their best fit value from Ref.~\cite{deSalas:2020pgw}, considering NO and $\sigma_k$ corresponding to the confidence level (CL) 68.27\% for each parameter $k$.

\section{Results and discussions}
\label{results}
In this section, we present our results on the projected limits and sensitivities to the isotropic $CPT-$odd SME coefficients $a_{\alpha \beta}$ with the experimental configuration of DUNE and the combination of DUNE and ESSnuSB analysis. For a comprehensive study of the projected sensitivities to the isotropic $CPT-$odd SME coefficients $a_{\alpha \beta}$ in the ESSnuSB proposal, see Ref.~\cite{Cordero:2024nho}.

In Figs.~\ref{f1dune} and \ref{f3dune}, we examine the correlations of the coefficients $a_{\alpha \beta}$ and $(a_{\alpha\alpha} - a_{\tau\tau})$ with $\delta_{CP}$. Moreover, we assess the impact of these coefficients on the future determination of $\delta_{CP}$. A robust measurement of $\delta_{CP}$ can be achieved even in the presence of Lorentz violating effects, as shown in Fig.~\ref{f6dune}. The correlations with $\sin^2 \theta_{23}$ are presented in Figs.~\ref{f2dune} and \ref{f4dune}. Just as importantly, the LIV phases, $\phi_{\alpha \beta}$, might compromise the determination of $\delta_{CP}$, and similarly for the atmospheric mixing angle, $\theta_{23}$~\cite{Pan:2023qln}. In Fig.~\ref{g1dune}, we show the projected sensitivities to the coefficients $a_{\alpha \beta}$ with respect to $\phi_{\alpha\beta}$ and $\Delta\chi^2$ for $a_{\alpha \alpha}$. Finally, in Fig.~\ref{g2dune}, we display the correlations of the LIV phases, $\phi_{\alpha \beta}$, with respect to $\delta_{CP}$.

\subsection{Correlations of the SME coefficients with $\delta_{CP}$}

\begin{figure}[H]
		\begin{subfigure}[h]{0.48\textwidth}
			\caption{  }
			\label{ff1}
\includegraphics[width=\textwidth]{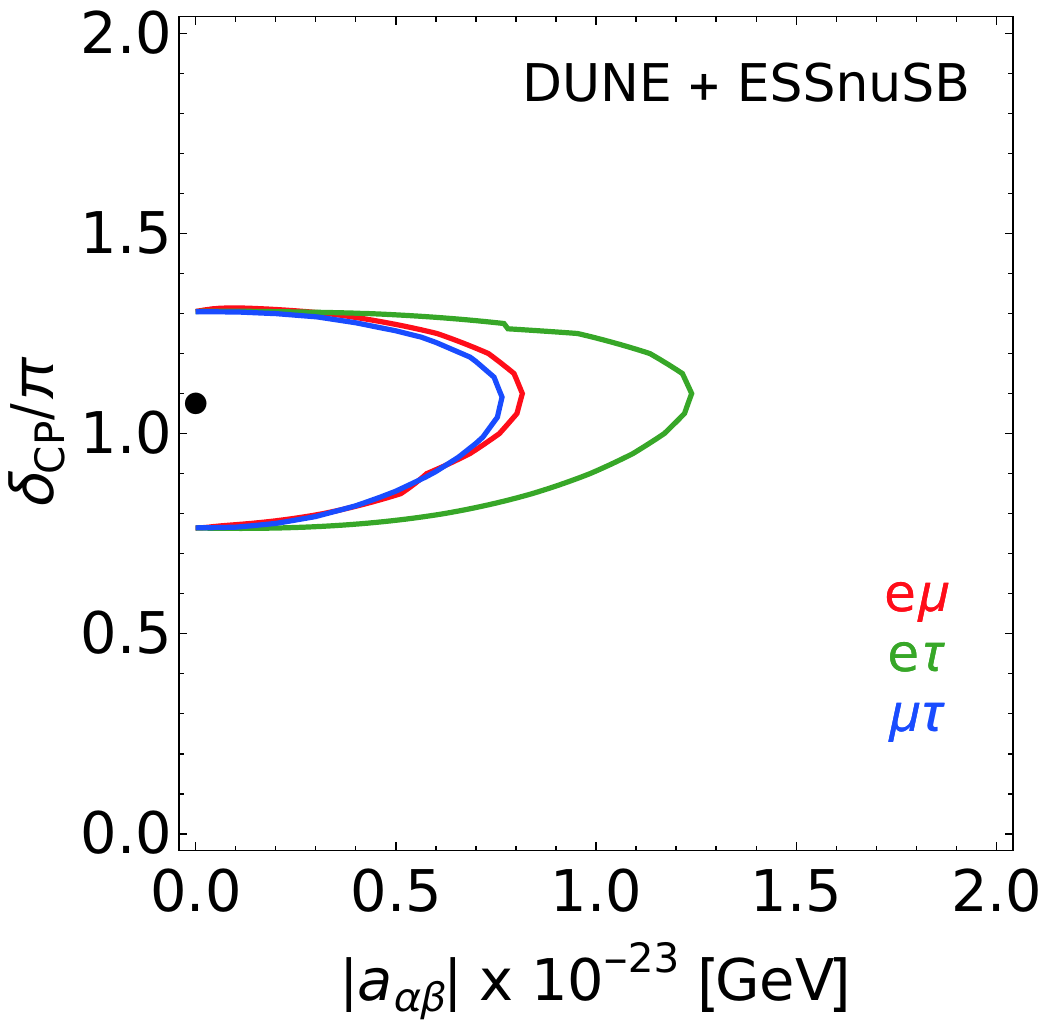}
		\end{subfigure}
		\hfill
		\begin{subfigure}[h]{0.48 \textwidth}
			\caption{}
			\label{ff2}
			\includegraphics[width=\textwidth]{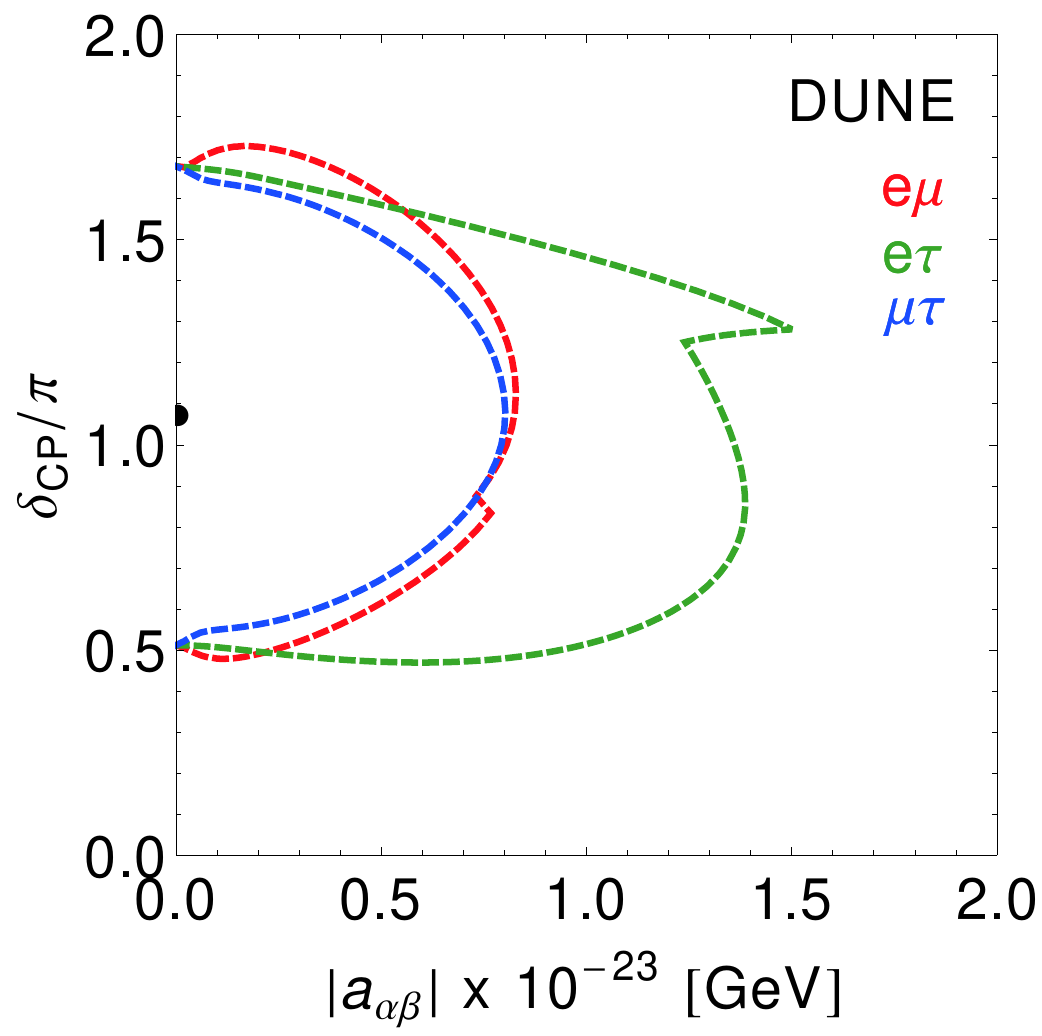}
		\end{subfigure}
		\hfill	
 \caption{Expected 95\% CL sensitivity regions in the $|a_{\alpha \beta}|-\delta_{CP}$ projection plane. The (red, green, blue)-dashed lines correspond to the SME coefficients from the $\alpha \beta = (e \mu, e \tau, \mu \tau)$ sectors, respectively. We have marginalized over $\theta_{23}$ around its 1$\sigma$ uncertainty~\cite{deSalas:2020pgw}, and corresponding LIV phases, $\phi_{\alpha \beta}$, free within [0$-2\pi$]. All of the remaining oscillation parameters are fixed to their NO best fit values. The DUNE simulated data analysis is displayed in the right panel, while the DUNE+ESSnuSB combined analysis is presented in the left panel. See the text for a detailed explanation.}
  \label{f1dune}
\end{figure}

The correlations among $\delta_{CP}$ and the off-diagonal SME coefficients, $|a_{\alpha \beta}|$, are shown in Fig.~\ref{f1dune}. In both situations, we have marginalized the atmospheric mixing angle, $\theta_{23}$, around its 1$\sigma$ uncertainty~\cite{deSalas:2020pgw}, as well as the $CPT-$odd phases, $\phi_{\alpha \beta}$, in the interval [0$-2\pi$]. For example, for the case of DUNE alone (right panel), we find that the presence of the flavor-changing parameters ($e-\mu$) and ($e-\tau$) has a greater impact on the determination of $\delta_{CP}$. However, the combination of DUNE and ESSnuSB (left panel) could provide a more robust determination of $\delta_{CP}$ within the aforementioned LIV scenario.

\begin{figure}[H]
		\begin{subfigure}[h]{0.46\textwidth}
			\caption{  }
			\label{ff6}
\includegraphics[width=\textwidth]{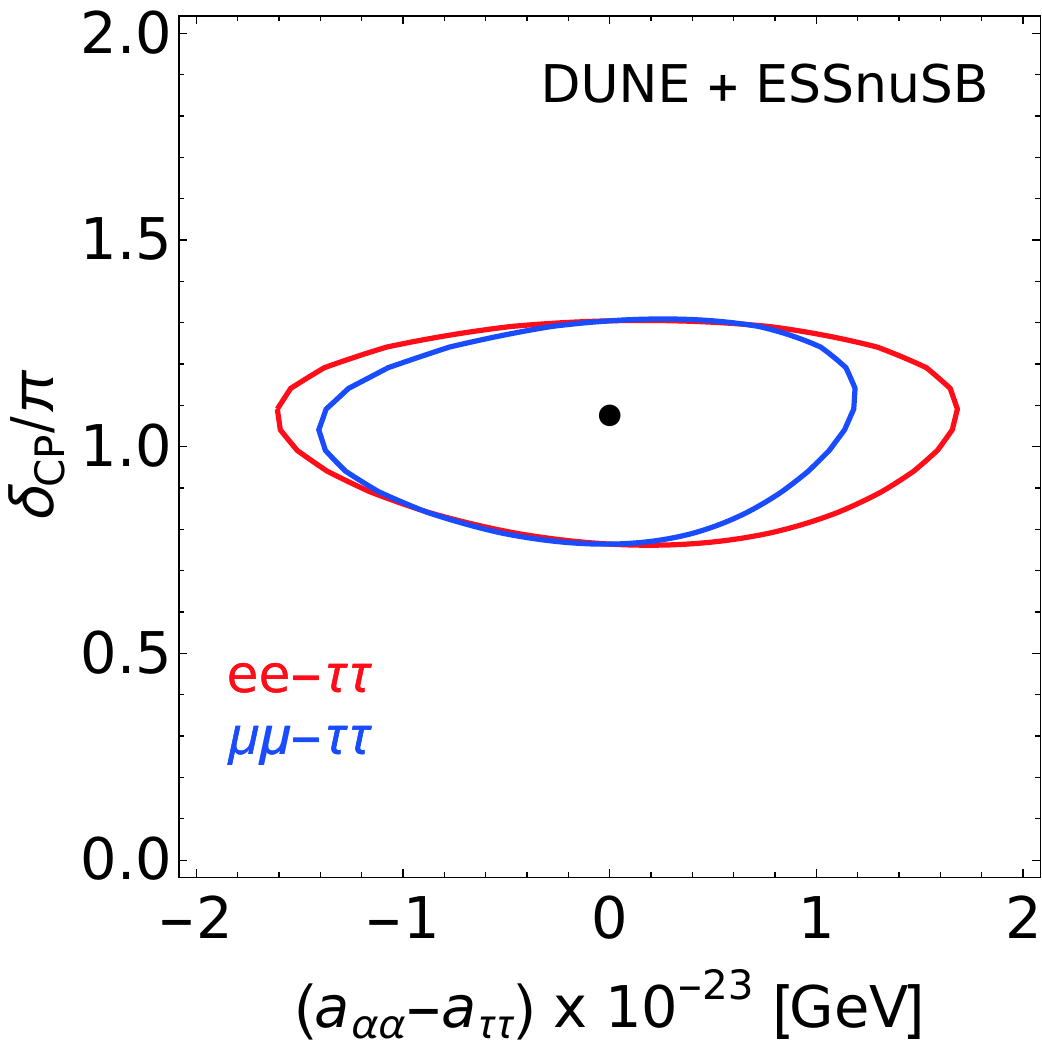}
		\end{subfigure}
		\hfill
		\begin{subfigure}[h]{0.47 \textwidth}
			\caption{}
			\label{ff7}
			\includegraphics[width=\textwidth]{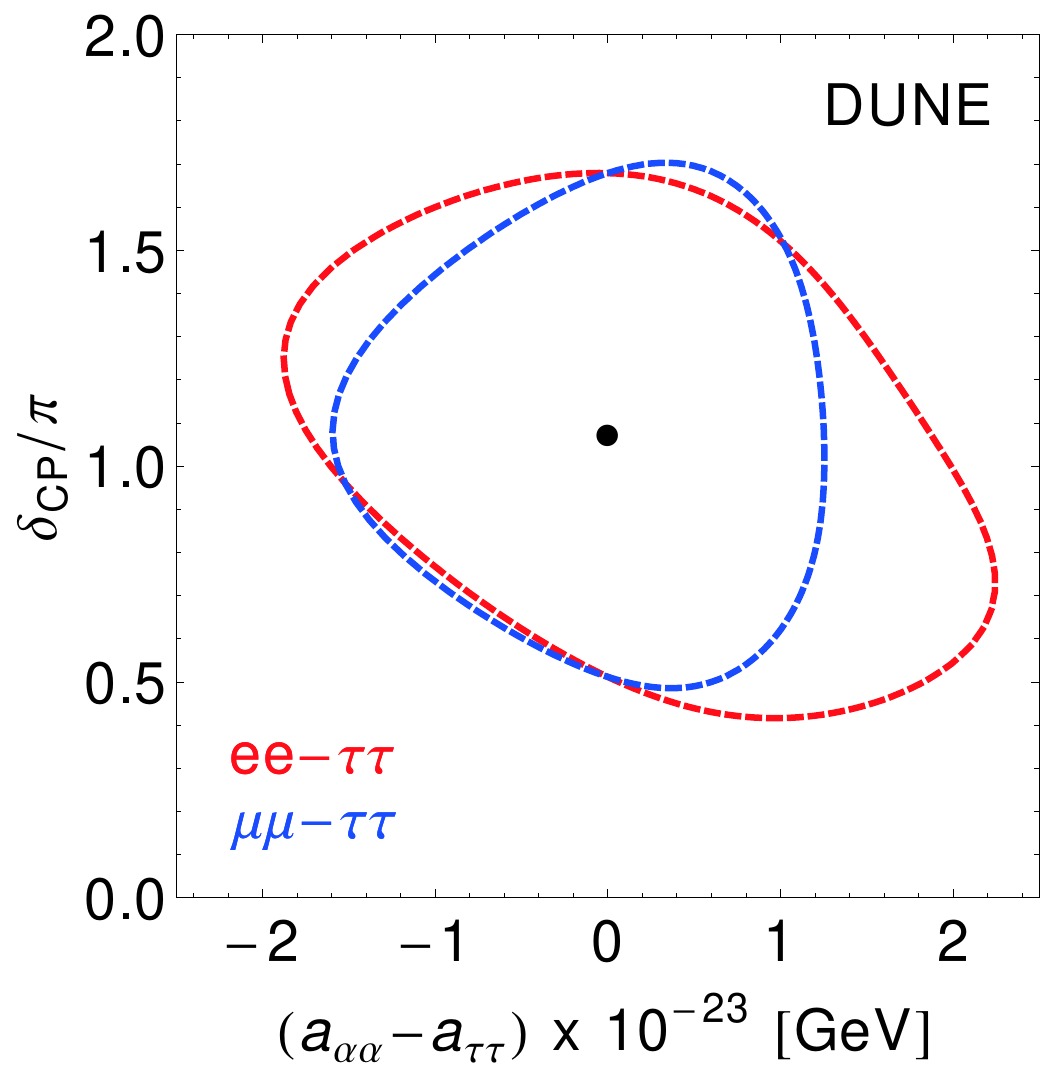}
		\end{subfigure}
		\hfill	
 \caption{Expected 95\% CL sensitivity regions in the $(a_{\alpha \alpha}-a_{\tau \tau})-\delta_{CP}$ projection plane. The (red, blue)-dashed lines correspond to $\alpha \alpha = (ee, \mu \mu)$, respectively. We have marginalized over $\theta_{23}$ around its 1$\sigma$ uncertainty~\cite{deSalas:2020pgw}. All of the remaining oscillation parameters are fixed to their NO best fit value. The DUNE simulated data analysis is shown in the right panel, while the DUNE+ESSnuSB combined analysis is displayed in the left panel. See the text for a detailed explanation.}
  \label{f3dune}
\end{figure}

Regarding diagonal SME coefficients, taking advantage of the freedom to redefine diagonal elements up to a global constant, we perform an analysis of the elements $a_{\alpha \alpha}-a_{\tau \tau}$. The correlations with respect to $\delta_{CP}$ and $\sin^2\theta_{23}$ are shown in Figs.~\ref{f3dune} and~\ref{f4dune}, accordingly. 

In Fig.~\ref{f3dune}, we display our sensitivity results corresponding to 95\% CL in the projection plane $(a_{\alpha \alpha}-a_{\tau \tau}) vs. \delta_{CP}$. We note that the determination of $\delta_{CP}$ is more affected by the linear combination $a_{ee}-a_{\tau \tau}$ than by $\mu\mu-\tau\tau$. Here, we have marginalized the atmospheric mixing angle $\theta_{23}$ around its 1$\sigma$ uncertainty~\cite{deSalas:2020pgw}. In addition, all of the remaining oscillation parameters were fixed to their best fit values from Table~\ref{tab:osc}. As in the case with non-diagonal SME coefficients (Fig.~\ref{f1dune}), under the presence of LIV from the diagonal sectors, the combination of DUNE and ESSnuSB (left panel of Fig.~\ref{f3dune}) may provide a more robust determination of $\delta_{CP}$ compared to DUNE alone (right panel of Fig.~\ref{f3dune}).
\begin{figure}[H]
\includegraphics[scale=0.5]{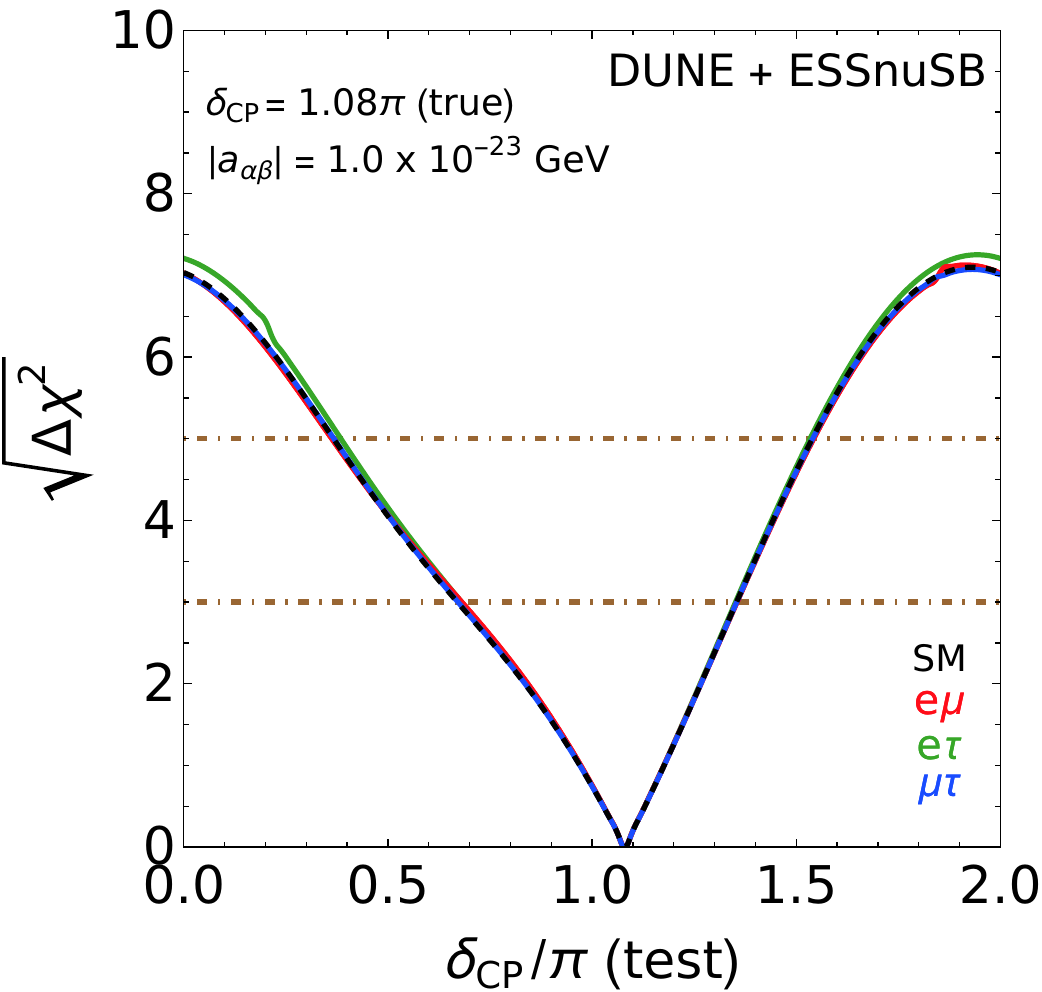}
		 \caption{The expected leptonic $CP$-phase precision sensitivity for the combination of DUNE and ESSnuSB. The (red, green, blue) lines correspond to the inclusion of non-zero SME coefficients from the $\alpha \beta = (e \mu, e \tau, \mu \tau)$ sectors, respectively. The black line represents the standard three neutrino oscillation picture (SM). The brown dot-dashed horizontal lines display the 3$\sigma$ and 5$\sigma$ CL, accordingly. We marginalize over $\theta_{23}$ around its 1$\sigma$ uncertainty~\cite{deSalas:2020pgw}, and corresponding LIV phases $\phi_{\alpha \beta}$ in the interval [0$-2\pi$]. The magnitude of the non-diagonal SME coefficients is fixed ($|a_{\alpha \beta}| = 1.0 \times 10^{-23}$ GeV). All of the remaining oscillation parameters are fixed to their NO best fit value. See the text for details.} 
  \label{f6dune}
\end{figure}
In Fig.~\ref{f6dune}, we show the precision sensitivity to the leptonic $CP$-phase in the presence of LIV. We observe that the precision sensitivity of $\delta_{CP}$ is slightly modified, with high significance ($\sqrt{\Delta \chi^2} \gtrsim 6\sigma$ CL), by including non-zero SME coefficients ($a_{\alpha \beta}$) from the off-diagonal ($e-\tau$) parameter, while remaining practically unchanged by the presence of parameters ($e-\mu$) and $(\mu-\tau)$. Hence, a robust measurement of $\delta_{CP}$ can be accomplished even in the presence of Lorentz violating effects from non-diagonal coefficients.

\subsection{Correlations of the SME coefficients with $\sin^2 \theta_{23}$}
\begin{figure}[H]
\begin{subfigure}[h]{0.48\textwidth}
			\caption{}
			\label{ff3}
\includegraphics[width=\textwidth]{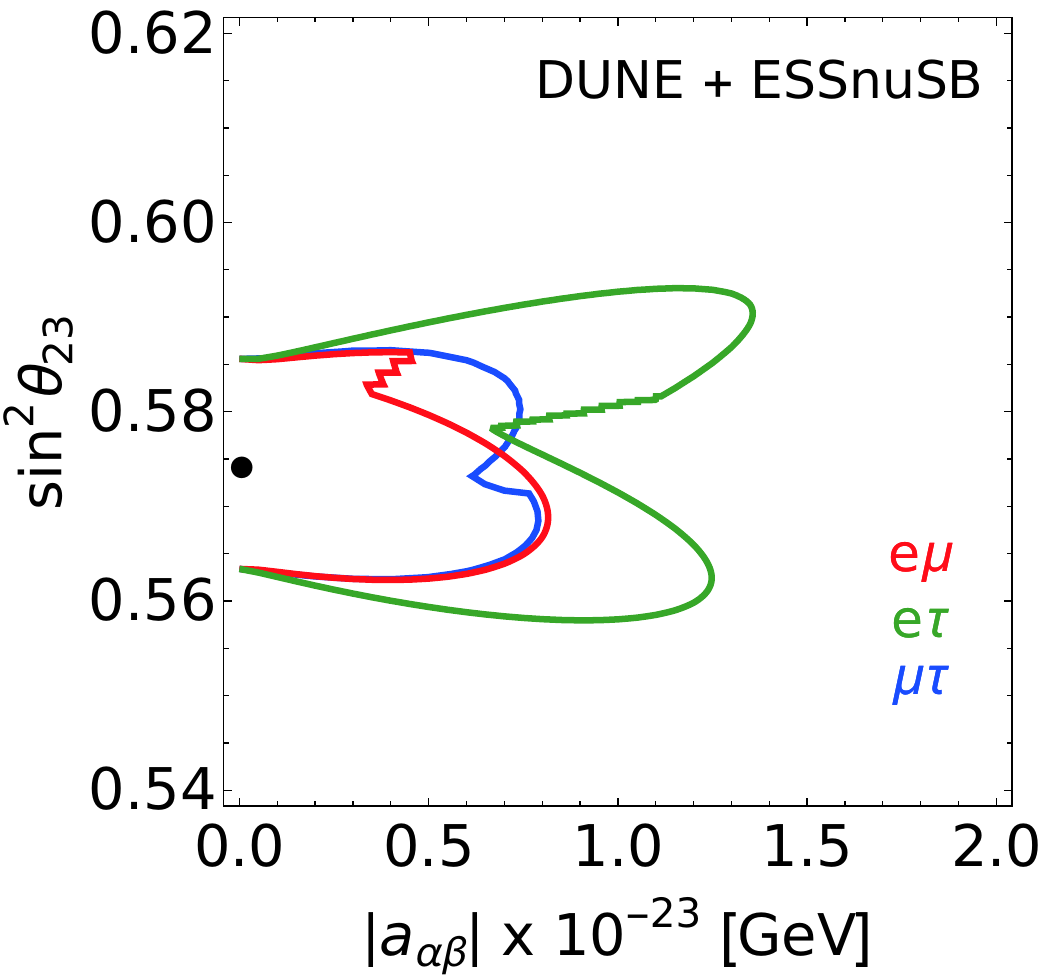}
		\end{subfigure}
		\hfill
		\begin{subfigure}[h]{0.465\textwidth}
			\caption{}
			\label{ff4}
			\includegraphics[width=\textwidth]{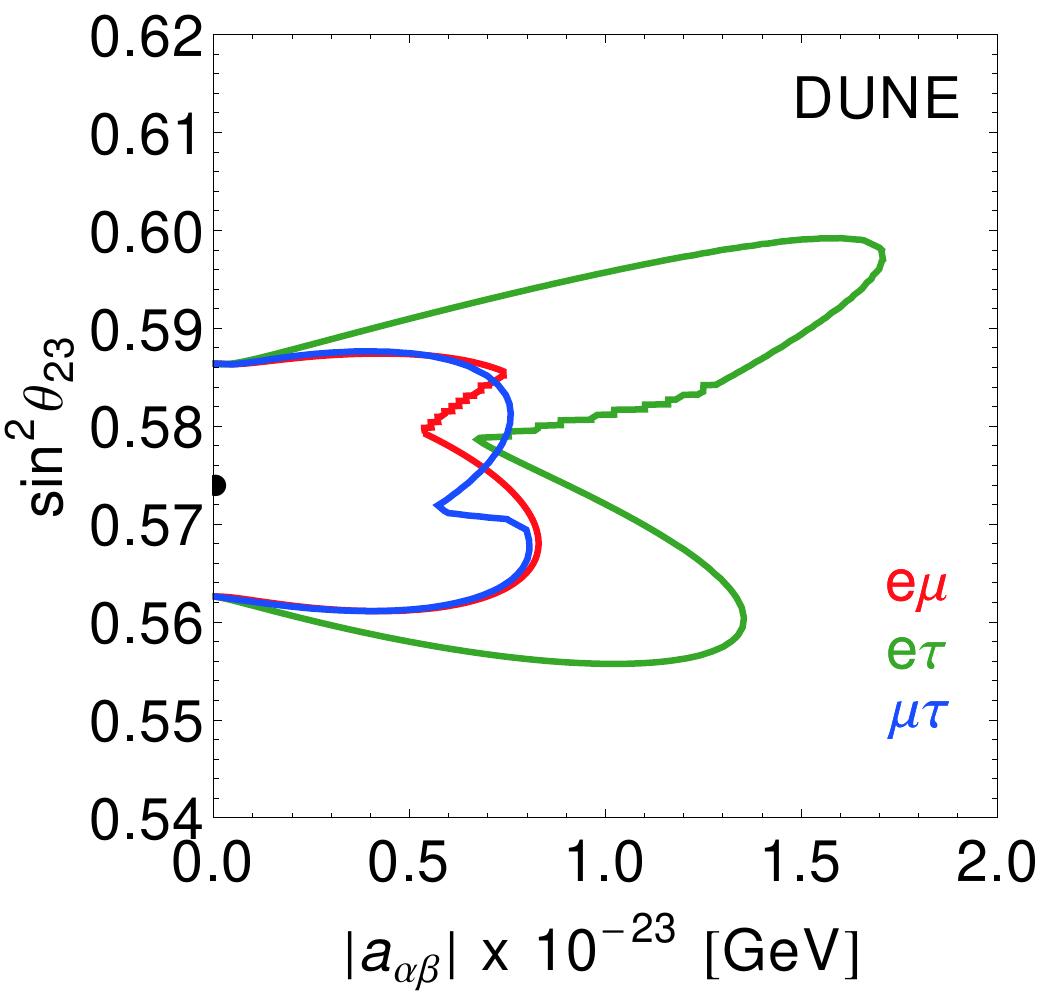}
		\end{subfigure}
		\hfill	
 \caption{Expected 95\% CL sensitivity regions in the $|a_{\alpha \beta}|-\sin^2\theta_{23}$ projection plane. The (red, green, blue)-solid lines correspond to the SME coefficients from the $\alpha \beta = (e \mu, e \tau, \mu \tau)$ sectors, respectively. We have marginalized over $\delta_{CP}$ around its 1$\sigma$ uncertainty~\cite{deSalas:2020pgw}, and corresponding LIV phases, $\phi_{\alpha \beta}$, from [0$-2\pi$]. All of the remaining oscillation parameters were fixed to their NO best fit values. The DUNE simulated data analysis is presented in the right panel, while the DUNE+ESSnuSB combined analysis is shown in the left panel. See the text for a detailed explanation.}
  \label{f2dune}
\end{figure}

In Fig.~\ref{f2dune}, we show the correlations between the off-diagonal SME coefficients, $|a_{\alpha \beta}|$, and the atmospheric mixing angle, $\sin^2\theta_{23}$. In this case, we marginalized over $\delta_{CP}$ around its 1$\sigma$ uncertainty~\cite{deSalas:2020pgw}, as well as the corresponding $CPT-$odd phases, $\phi_{\alpha \beta}$, in the interval [0$-2\pi$]. In both cases, we note that the determination of $\sin^2\theta_{23}$ is primarily impacted by the SME coefficients $a_{e \tau}$, while the ($e-\mu$) and ($\mu-\tau$) parameters have a mild influence on the determination of the atmospheric mixing angle. In contrast to the case of $\delta_{CP}$, the combination of DUNE and ESSnuSB (left panel of Fig.~\ref{f2dune}) could only have a mild enhancement in the determination of $\sin^2 \theta_{23}$. However, the expected limit set on the non-diagonal $e-\tau$ coefficient could be significantly improved with respect to the scenario with DUNE alone (right panel of Fig.~\ref{f2dune}).

\begin{figure}[H]
		\begin{subfigure}[h]{0.47\textwidth}
			\caption{  }
			\label{ff8}
\includegraphics[width=\textwidth]{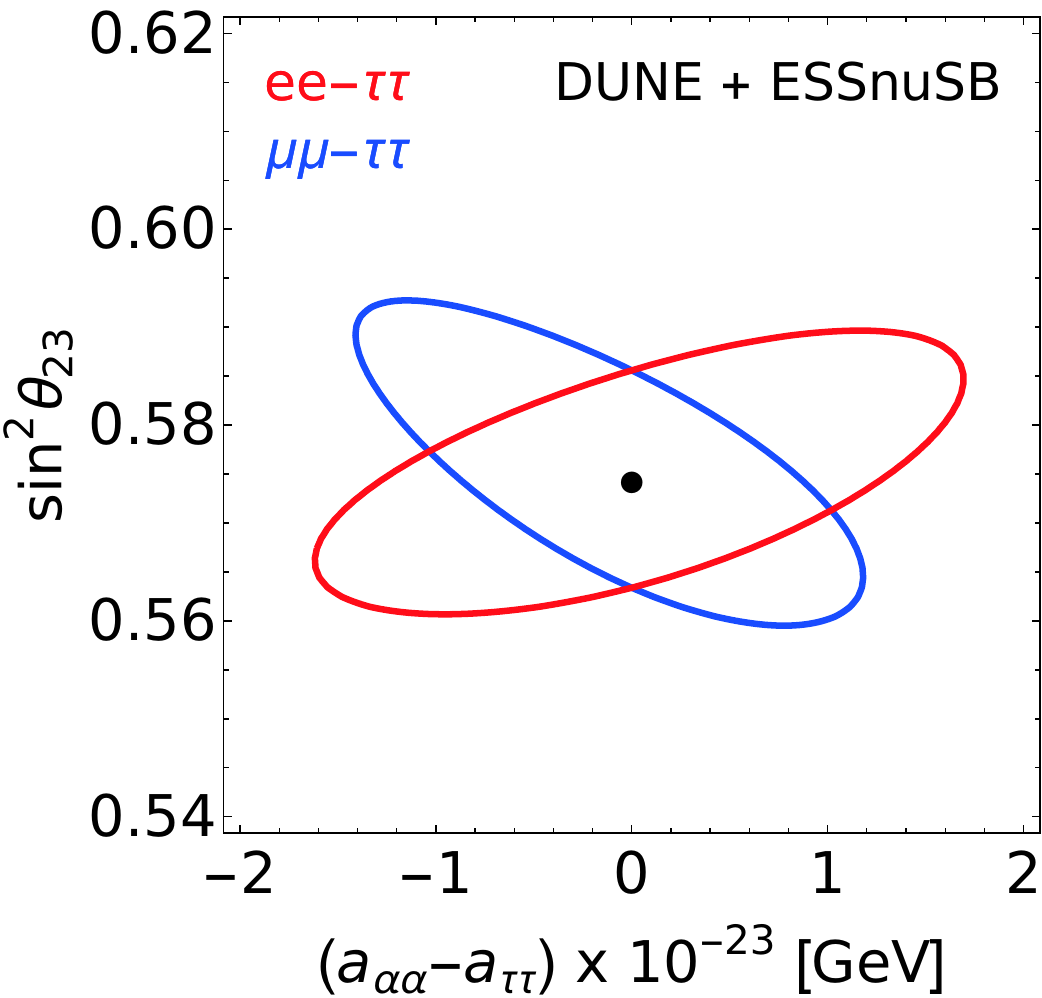}
		\end{subfigure}
		\hfill
		\begin{subfigure}[h]{0.47\textwidth}
			\caption{}
			\label{ff9}
			\includegraphics[width=\textwidth]{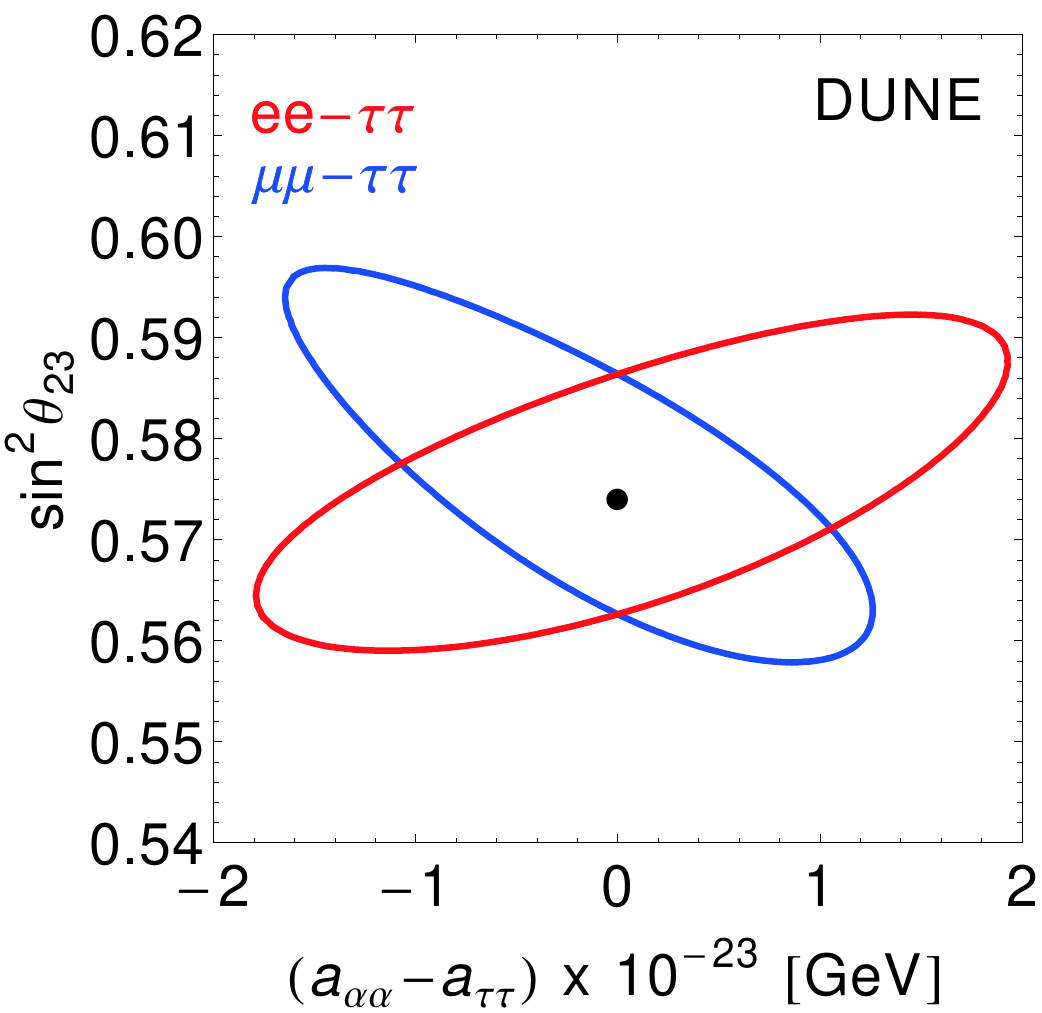}
		\end{subfigure}
		\hfill	
 \caption{Expected 95\% CL sensitivity regions in the $a_{\alpha \alpha}-\sin^2\theta_{23}$ projection planes. The (red, blue)-solid lines correspond to $\alpha \alpha = (ee, \mu \mu)$, respectively. We have marginalized over $\delta_{CP}$ around its 1$\sigma$ uncertainty~\cite{deSalas:2020pgw}. All of the remaining oscillation parameters are fixed to their NO best fit value. The DUNE simulated data analysis is presented in the right panel, while the DUNE+ESSnuSB combined analysis is shown in the left panel. See the text for a detailed explanation.}
  \label{f4dune}
\end{figure}

In Fig.~\ref{f4dune}, we observe that the diagonal SME coefficients have a similar impact on the determination of the mixing angle, showing a positive (negative) correlation for the isotropic SME coefficients, $a_{ee}-a_{\tau \tau}$ ($a_{\mu \mu}-a_{\tau \tau}$). In a similar fashion to the non-diagonal SME coefficients (Fig.~\ref{f2dune}), under the presence of LIV from the diagonal sectors, the combination of DUNE and ESSnuSB (left panel of Fig.~\ref{f4dune}) may only provide a slight improvement in the determination of $\sin^2\theta_{23}$ compared to DUNE alone (right panel of Fig.~\ref{f4dune}).

\subsection{SME parameter sensitivities and correlations among LIV phases ($\phi_{\alpha \beta}$) with $\delta_{CP}$}
\begin{figure}[H]
		\begin{subfigure}[h]{0.47\textwidth}
			\caption{  }
			\label{gg1}
\includegraphics[width=\textwidth]{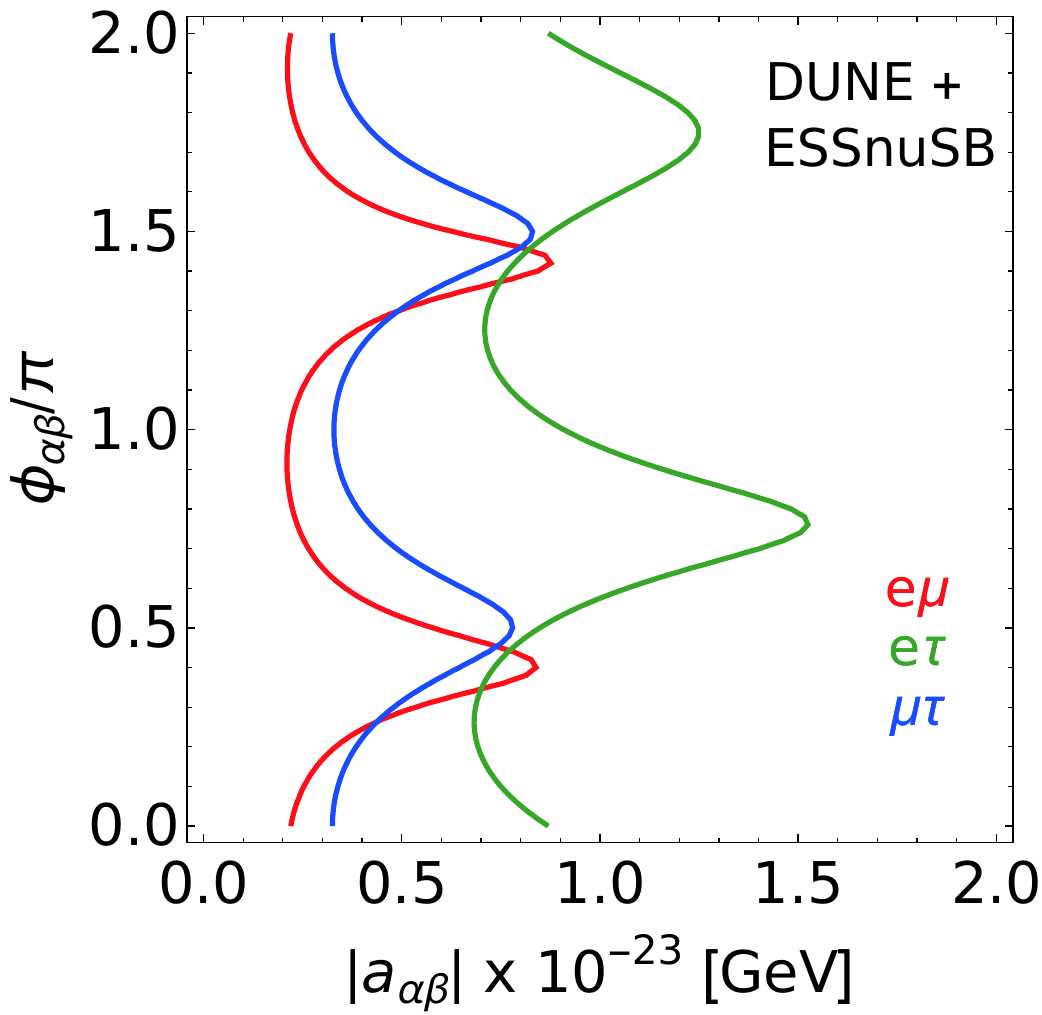}
		\end{subfigure}
		\hfill
		\begin{subfigure}[h]{0.46 \textwidth}
			\caption{}
			\label{gg2}
			\includegraphics[width=\textwidth]{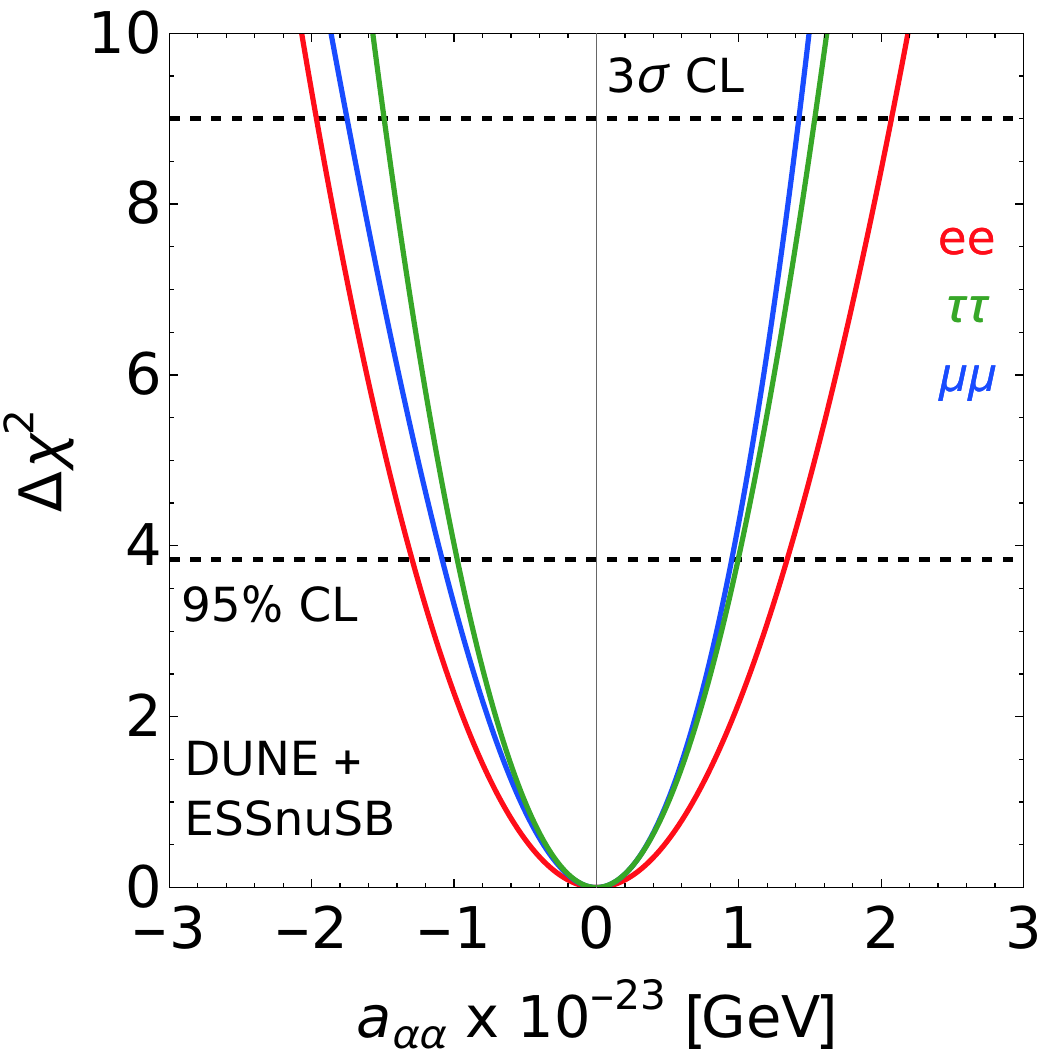}
		\end{subfigure}
		\hfill	
 \caption{Expected 95\% CL sensitivity regions in the $\phi_{\alpha \beta}$ vs. $|a_{\alpha \beta}|$ projection planes (left panel) and expected sensitivities to the diagonal SME coefficients: $a_{\alpha \alpha}$ (right panel). The (red, green, blue)-dashed lines on the left correspond to the SME coefficients from the $\alpha \beta = (e \mu, e \tau, \mu \tau)$ sectors, respectively. The colored lines on the right correspond to the diagonal elements. We marginalize over $\theta_{23}$ and $\delta_{CP}$ around their 1$\sigma$ uncertainty~\cite{deSalas:2020pgw}. All of the remaining oscillation parameters are fixed to their NO best fit value. See the text for a detailed explanation.}
  \label{g1dune}
\end{figure}

In the left panel of Fig.~\ref{g1dune}, we present our results of the expected sensitivities to the SME coefficients $a_{\alpha \beta}$ for the combination of DUNE and ESSnuSB, the projected 95\% CL sensitivities to the non-diagonal isotropic SME coefficients are $|a_{e \tau}|\lesssim 1.6 \times 10^{-23}$ GeV, $|a_{e \mu}|\lesssim 0.9 \times 10^{-23}$ GeV, and $|a_{\mu \tau}|\lesssim 0.8 \times 10^{-23}$ GeV, accordingly. However, existing bounds on the isotropic SME coefficients from the Super-Kamiokande experiment set $|a_{e \tau}| < 2.8 \times 10^{-23}$ GeV, $|a_{e \mu}| < 1.8 \times 10^{-23}$ GeV, and $|a_{ \mu \tau}| < 5.1 \times 10^{-24}$ GeV~\cite{Super-Kamiokande:2014exs}; all bounds at 95\% CL. Hence, the combination of DUNE and ESSnuSB could improve the limits set on the non-diagonal isotropic SME coefficients $|a_{e \mu}|$ and $|a_{e \tau}|$ with respect to those from Super-Kamiokande. Furthermore, the right panel of Fig.~\ref{g1dune} displays the sensitivities to the diagonal SME coefficients $a_{\alpha \alpha}$, the projected 95\% CL sensitivities are: $a_{ee} \in (-1.3, \, 1.4) \times 10^{-23}$ GeV, $a_{\mu \mu} \in (-1.1, \, 0.9) \times 10^{-23}$ GeV, and $a_{\tau \tau} \in (-0.9, \, 0.9) \times 10^{-23}$ GeV, respectively. Compared with the combination of T2HK, DUNE and ICAL, as well as other existing limits in the literature (see Table~III and Fig.~(10) of Ref.~\cite{Raikwal:2023lzk}), the combination of DUNE and ESSnuSB could set the strongest limits on the coefficient $a_{\mu \mu}$. However, current bounds from the IceCube collaboration~\cite{IceCube:2021tdn} on the $a_{ee}$ and $a_{\tau \tau}$ coefficients are several orders of magnitude stronger than the expected limits from the combination of DUNE and ESSnuSB experiments.

\begin{figure}[H]
		\begin{subfigure}[h]{0.48\textwidth}
			\caption{  }
			\label{gg3}
\includegraphics[width=\textwidth]{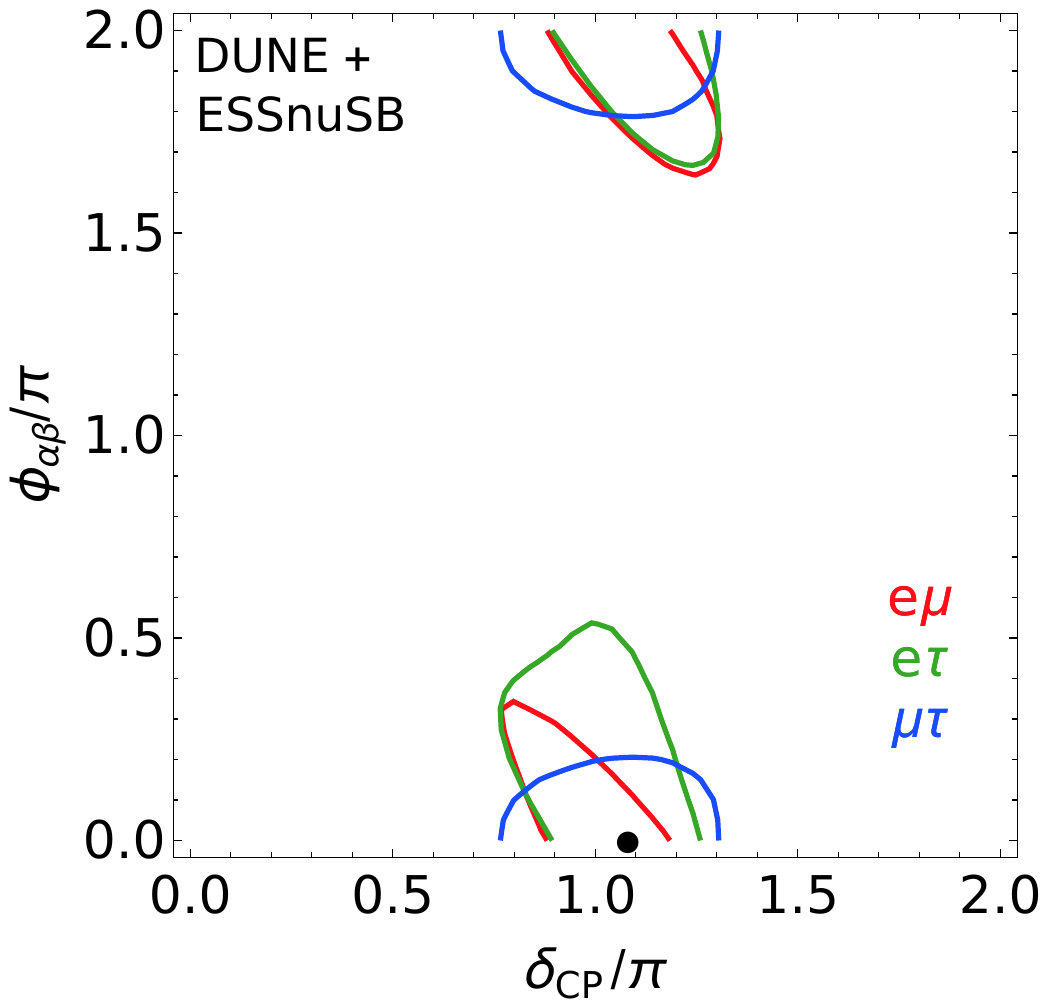}
		\end{subfigure}
		\hfill
		\begin{subfigure}[h]{0.48 \textwidth}
			\caption{}
			\label{gg4}
			\includegraphics[width=\textwidth]{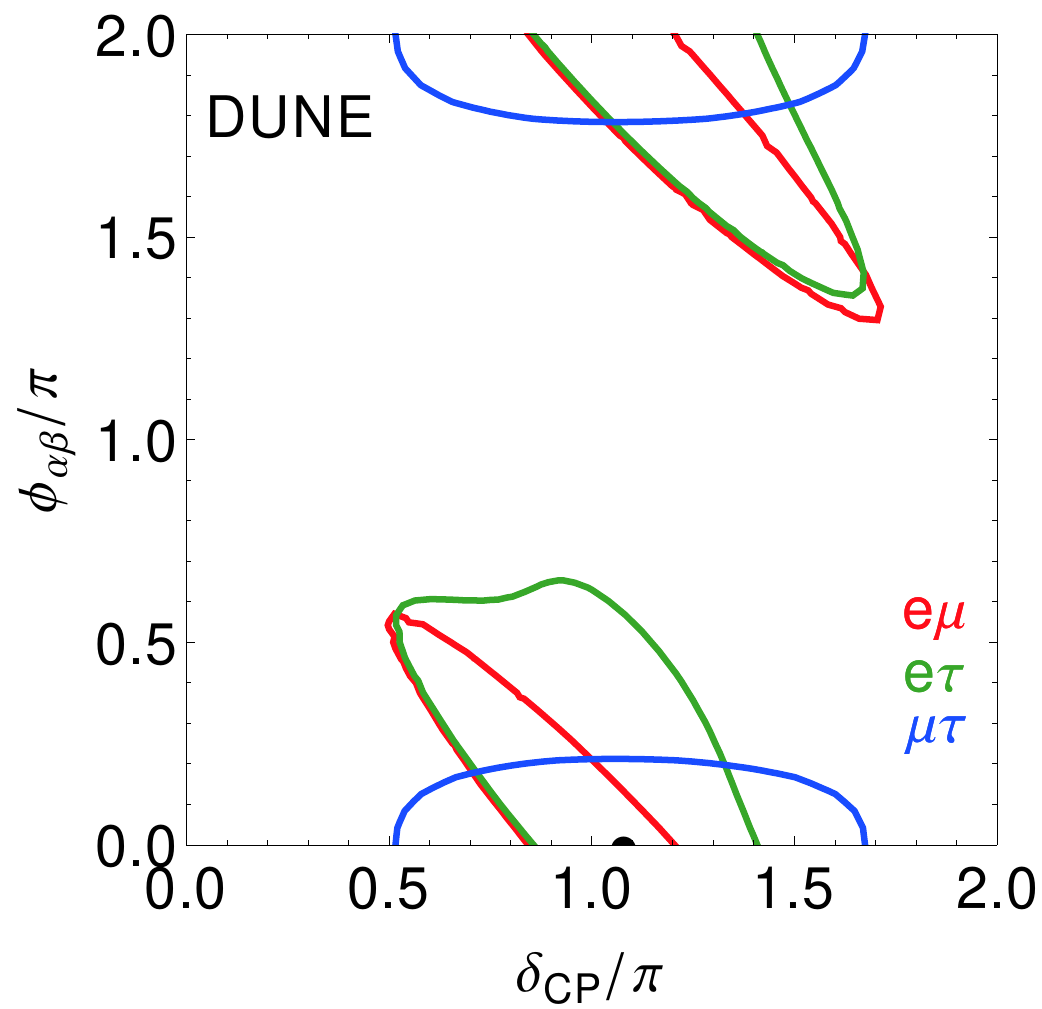}
		\end{subfigure}
		\hfill	
 \caption{Expected 95\% CL sensitivity regions in the $\phi_{\alpha \beta}-\delta_{CP}$ projection plane. The (red, green, blue)-dashed lines correspond to the LIV phases from the $\alpha \beta = (e \mu, e \tau, \mu \tau)$ sectors, respectively. We marginalize over $\theta_{23}$ around its 1$\sigma$ uncertainty~\cite{deSalas:2020pgw}. The magnitude of the non-diagonal SME coefficients is fixed ($|a_{\alpha \beta}| = 1.0 \times 10^{-23}$ GeV). All of the remaining oscillation parameters were fixed to their NO best fit value. The DUNE simulated data analysis is shown in the right panel, while the DUNE+ESSnuSB combined analysis is presented in the left panel. See the text for a detailed explanation.}
  \label{g2dune}
\end{figure}

In Fig.~\ref{g2dune}, we display the correlations of $\delta_{CP}$ with the corresponding LIV phases $\phi_{\alpha \beta}$. The magnitude of the non-diagonal SME coefficients was fixed to $|a_{\alpha \beta}| = 1.0 \times 10^{-23}$ GeV. We notice that the LIV phases, $\phi_{\alpha \beta}$, have a similar impact on the reconstructed $\delta_{CP}$ phase. Furthermore, the expected 95\% CL allowed values of the leptonic $CP$ phase are: $\delta_{CP} / \pi= 1.08 \pm 0.6$ (DUNE alone) and $\delta_{CP} / \pi= 1.08^{+0.22}_{-0.3}$ (DUNE and ESSnuSB). The coefficients $a_{e\mu}$ and $a_{e\tau}$ show higher correlations.

\section{Conclusions}
\label{conclusions}

High-energy neutrino oscillations provide an excellent opportunity to investigate potential breakdowns of the Lorentz and $CPT$ symmetries.
In this study, we have analyzed the impact of isotropic $CPT-$odd Lorentz violation on the determination of the mixing angle $\theta_{23}$ and the leptonic $CP$ phase $\delta_{CP}$, as well as projected sensitivities and correlations, considering a DUNE-like setup and the combined DUNE-ESSnuSB analysis.

The determination of the atmospheric mixing angle $\theta_{23}$ and phase $\delta_{CP}$ will be mainly affected by the effects of the isotropic $CPT-$odd SME coefficients from the ($e-\tau$) sector; effects from the ($e-\mu$) sector are moderate, and the impact from the ($\mu-\tau$) sector is muted: Figs.~\ref{f1dune} and~\ref{f2dune}. Effects from the diagonal ($ee-\tau \tau$) sector will have a considerable impact on the determination of the leptonic phase $CP$ (Fig.~\ref{f3dune}). Moreover, we have shown that by performing a combined fit of the DUNE and ESSnuSB datasets, the determination of $\delta_{CP}$ remains robust against LIV effects from the non-diagonal coefficients (Fig.~\ref{f6dune}). In addition, we observed that the presence of the corresponding LIV phases, $\phi_{\alpha \beta}$, may have a similar effect on the reconstructed $CP$ phase, showing a significant correlation among flavor-changing coefficients, ($e-\mu$) and ($e-\tau$), accordingly (Fig.~\ref{g2dune}). Regarding diagonal elements, ($ee-\tau \tau$) and ($\mu \mu-\tau \tau$), the determination of the atmospheric mixing angle $\theta_{23}$ could be slightly influenced by the presence of the isotropic SME coefficients, the combination of a DUNE and ESSnuSB fit may only provide a marginal improvement in the determination of $\sin^2\theta_{23}$ relative to the case of a DUNE-only fit (Fig.~\ref{f4dune}).

On the other hand, for the combination of DUNE and ESSnuSB, the projected 95\% CL sensitivities to the non-diagonal isotropic SME coefficients are $|a_{e \tau}|\lesssim 1.6 \times 10^{-23}$ GeV, $|a_{e \mu}|\lesssim 0.9 \times 10^{-23}$ GeV, and $|a_{\mu \tau}|\lesssim 0.8 \times 10^{-23}$ GeV, accordingly (left panel of Fig.~\ref{g1dune}). Furthermore, the projected 95\% CL sensitivities to the diagonal SME coefficients $a_{\alpha \alpha}$ are $a_{ee} \in (-1.3, \, 1.4) \times 10^{-23}$ GeV, $a_{\mu \mu} \in (-1.1, \, 0.9) \times 10^{-23}$ GeV, and $a_{\tau \tau} \in (-0.9, \, 0.9) \times 10^{-23}$ GeV, respectively (right panel of Fig.~\ref{g1dune}).

Although current limits set on the ($e-\tau$) and ($e-\mu$) coefficients from the Super-Kamiokande experiment can be improved considering the combination of the future DUNE and ESSnuSB configurations, this may not be the case for the ($\mu-\tau$) coefficient, which confirms a complementarity between the aforementioned experiments.

\section*{Acknowledgments}
We would like to acknowledge J.~S.~Diaz for useful discussions. We thank the anonymous referee for the comments and suggestions that have helped us to improve our manuscript. This work was partially supported by SNII-M\'exico and CONAHCyT research Grant No.~A1-S-23238.

This paper represents the views of the authors and should not be considered a DUNE or ESSnuSB collaboration paper.

\appendix

\section{Correlations of time and $Z-$spatial components of the $CPT-$odd SME coefficients}
\label{angles}

Long-baseline neutrino experiments have sensitivity to $CPT$ symmetry and Lorentz invariance violations through SME coefficients. This sensitivity is proportional to their baseline. One can see this through the oscillation probability expression, which can be parameterized as~\cite{Diaz:2009qk}
\begin{equation}
\label{eq:lbl-prob}
    P_{\nu_\beta \to \nu_\alpha}^{(1)} = 2L(P_\mathcal{C}^{(1)})_{\alpha\beta}\,,
\end{equation}
where $L$ is the baseline and
\begin{equation}\label{eq:pc}
 (P_\mathcal{C}^{(1)})_{\alpha\beta}={\rm Im}((S^{(0)}_{\alpha\beta})^*(\mathcal{C}^{(1)})_{\alpha\beta})\,.
\end{equation}
In Eq.~\eqref{eq:pc},
\begin{equation}
    |(S^{(0)}_{\alpha\beta})|^2 = P^{(0)}_{\nu_\beta\to\nu_\alpha}
\end{equation}
is the standard three neutrino flavor oscillation probability. Considering only the $CPT-$odd coefficients $(\Tilde{a}_L)^{A}_{\alpha\beta}$, for $A=T,Z$ and $\alpha,\beta=e,\mu,\tau$,
\begin{equation}
    (\mathcal{C}^{(1)})_{\alpha\beta} = (\Tilde{a}_L)^T_{\alpha\beta} - \hat{N}^Z (\Tilde{a}_L)^Z_{\alpha\beta}\,,
    \label{eq:c1}
\end{equation}
where
\begin{equation}
\label{eq:nz}
    \hat{N}^Z = -\sin\chi\sin\theta\cos\phi+\cos\chi\cos\theta,
\end{equation}
is the $Z$ component of the vector representing the neutrino propagation direction in the Sun-centered frame in terms of local spherical coordinates at the detector (see Fig.~2 of Ref.~\cite{Bailey:2006fd} for an illustration). In Eq.~\eqref{eq:nz}, $\chi$ is the detector colatitude, $\theta$ is the angle at the detector between the beam direction and vertical, and $\phi$ the angle between the beam and east of south. Notice that we are not taking into account sidereal variations so the components $A=X,Y$ of $(\Tilde{a}_L)^{A}_{\alpha\beta}$ are not relevant.

In recent neutrino oscillation analyzes~\cite{Fiza:2022xfw,Agarwalla:2023wft,Majhi:2022fed,Sarker:2023mlz}, the possible violation of $CPT$ symmetry, which would imply violation of the Lorentz invariance principle, was studied in the framework of the SME, in such a way that $CPT-$odd coefficients can have their time component constrained through neutrino propagation in long baseline experiments. However, from equations \eqref{eq:lbl-prob}-\eqref{eq:c1}, we see that if the $Z-$component of the same coefficient is considered in the analysis, a correlation arises between them, which affects the predicted limits on the time component alone.

Now, we shift our attention to the Hamiltonian picture, for which the relevant coefficients are $(a_L)^A_{\alpha \beta}$, without approximation. Explicitly, considering only the relevant $CPT-$odd coefficients of the SME neutrino sector:
\begin{equation}
\label{LIVHAM}
 \hat{H}_{\text{LIV}} =   \left(
\begin{array}{ccc}
 a_{ee} & a_{e \mu} & a_{e \tau} \\
 a_{e \mu}^{*} & a_{\mu \mu} &  a_{ \mu \tau} \\
 a_{e \tau}^{*} & a_{ \mu \tau}^{*} & a_{ \tau \tau} \\
\end{array}
\right) - \hat{N}^Z \left(
\begin{array}{ccc}
 a_{ee}^Z & a_{e \mu}^Z & a_{e \tau}^Z \\
 a_{e \mu}^{Z*} & a_{\mu \mu}^Z &  a_{ \mu \tau}^Z \\
 a_{e \tau}^{Z*} & a_{ \mu \tau}^{Z*} & a_{ \tau \tau}^Z \\
\end{array}
\right)\,,
\end{equation}
where $a_{\alpha\beta}$ represent the time component and $a_{\alpha\beta}^Z$ the $Z-$spatial component of the coefficients for each neutrino oscillation sector ($\alpha,\beta=e,\mu,\tau\,$).

In our analysis, we consider a fixed value of $\hat{N}^Z$. Here we show how the value of $\hat{N}^Z$ changes in accordance with the determination of the experiment location.

\begin{figure}[H]
\begin{subfigure}[h]{0.48\textwidth}
			\caption{  }
			\label{ap:ang:left}
\includegraphics[width=\textwidth]{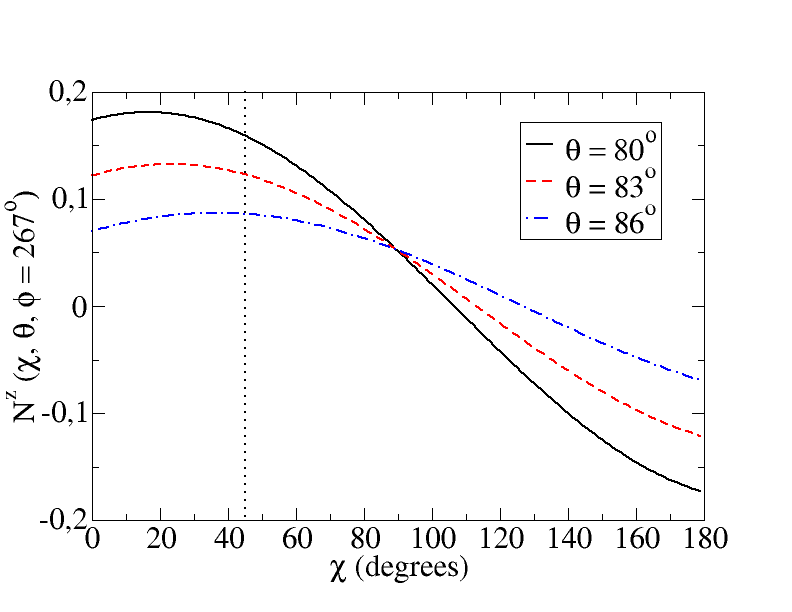}
		\end{subfigure}
		\hfill
		\begin{subfigure}[h]{0.48 \textwidth}
			\caption{}
			\label{ap:ang:right}
			\includegraphics[width=\textwidth]{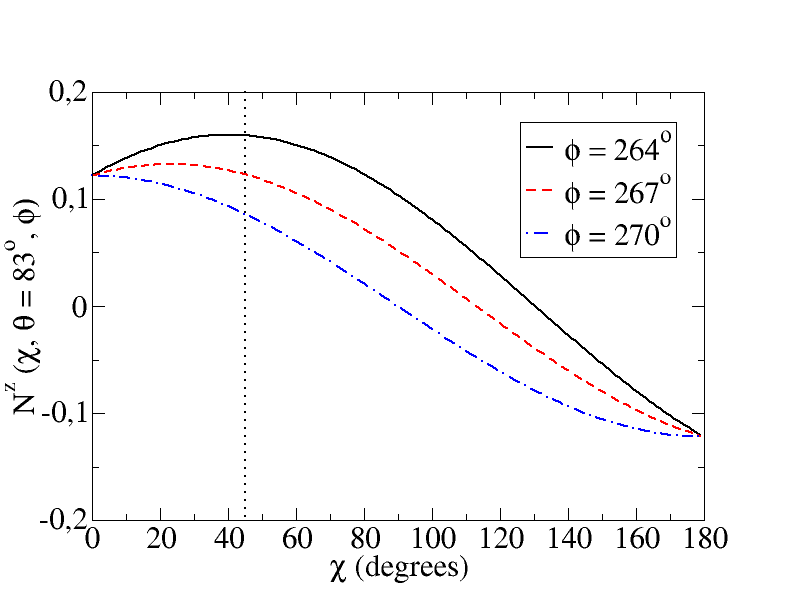}
		\end{subfigure}
		\hfill	
		 \caption{Values of $\hat{N}^Z$ as a function of $\chi$ in degrees. On the left side, $\phi=267^\circ$ and $\theta=80^\circ,83^\circ,$ and $86^\circ$ for black, dashed red, and dot-dashed blue lines, respectively. On the right side, $\theta=83^\circ$ and $\phi=264^\circ,267^\circ,$ and $270^\circ$ for black, dashed red, and dot-dashed blue lines, respectively. The vertical dotted line indicates the value of $\chi=45.6^\circ$, used in our analysis.}
  \label{ap:angles}
\end{figure}

\begin{table}[ht]
\caption{DUNE beam direction in terms of the Sun-centered frame angle definitions.}
  \label{tab:1}
    \begin{tabular}{c c c}
    \hline \hline
    Parameter & Definition & Value   \\ 
    \hline 
    $\chi$   & Colatitude & $ 45.6^{\circ} $\\   
    $\theta$ & Angle between beam and vertical & $ 84.2^{\circ} $\\   
    $\phi$  & Angle between beam and east of south & $262.5^{\circ} $\\ 
    \hline \hline  
    \end{tabular}
\end{table}

As shown in Fig.~\ref{ap:angles}, and from Eqs.~(\ref{eq:c1}) and (\ref{eq:nz}), experiment location resulting in bigger values of $\hat{N}^Z$ could enhance the sensitivity of the $Z-$spatial SME coefficients $(\tilde{a}_L)^Z_{\alpha \beta}$.

\begin{figure}[H]
\begin{subfigure}[h]{0.48\textwidth}
			\caption{  }
			\label{ff5a}
\includegraphics[width=\textwidth]{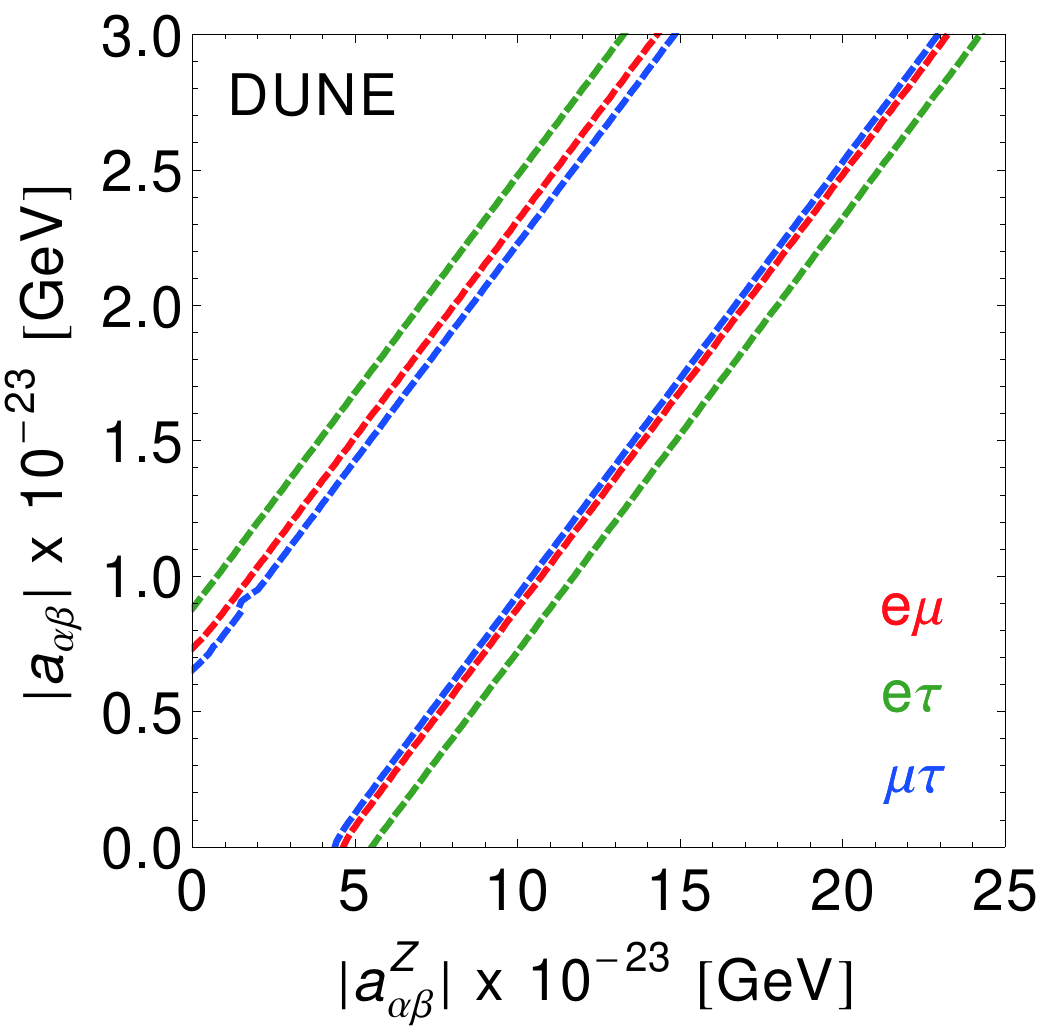}
		\end{subfigure}
		\hfill
		\begin{subfigure}[h]{0.48 \textwidth}
			\caption{}
			\label{ff5b}
			\includegraphics[width=\textwidth]{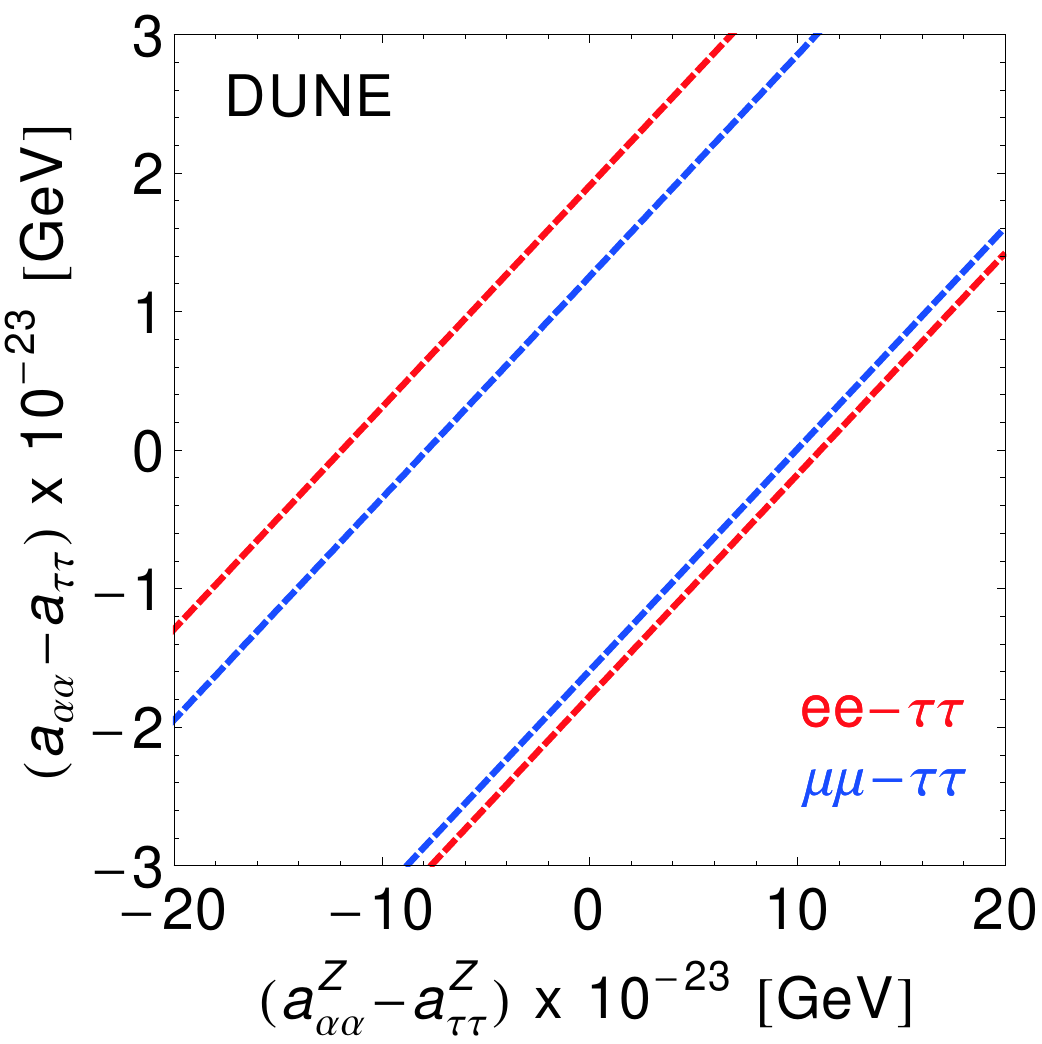}
		\end{subfigure}
		\hfill	
		 \caption{Expected 95\% CL sensitivity regions in the $|a_{\alpha \beta}|$ vs $|a_{\alpha \beta}^Z|$ and $(a_{\alpha \alpha}-a_{\tau \tau} )$ vs $(a_{\alpha \alpha}^Z-a_{\tau \tau}^Z)$ projection planes, respectively. In the left panel, the (red, green, blue)-dashed lines correspond to the SME coefficients from the $\alpha \beta = (e \mu, e \tau, \mu \tau)$ sectors. We have marginalized over the corresponding LIV phases, $\phi_{\alpha \beta}$ and $\phi_{\alpha \beta}^Z$, in the interval [0$-2\pi$]. In the right panel, $\alpha \alpha = (ee, \mu \mu)$, respectively, for red and blue colors. We marginalize over $\theta_{23}$ and $\delta_{CP}$ around their 1$\sigma$ uncertainty~\cite{deSalas:2020pgw}. All of the remaining oscillation parameters are fixed to their NO best fit values. See the text for a detailed explanation.}
  \label{f5dune}
\end{figure}

Previous evaluations on the isotropic SME coefficients $a_{\alpha \beta}$ considered $a_{\alpha \beta}^Z=0$~\cite{Barenboim:2018ctx, Sarker:2023mlz, Raikwal:2023lzk, Agarwalla:2023wft, Pan:2023qln}. Here, as an illustration, we consider the possibility of correlations between $a_{\alpha \beta}$ and $a_{\alpha \beta}^Z$ in the DUNE configuration. As far as the ESSnuSB proposal is concerned, since the precise location of the detector is still unknown, we will not consider the coefficients $a_{\alpha \beta}^Z$ for this configuration.
The correlation between time and the $Z-$component of the $CPT-$odd coefficients for DUNE is depicted in Fig.~\ref{f5dune}. Because the $Z-$component is proportional to $\hat{N}^Z$, which is relatively small for the DUNE location ($\hat{N}^Z \simeq 0.16\,$), constraints on the time component are stronger in this case. We obtain $\hat{N}^Z$ from the angle values relative to the DUNE location found in Table~\ref{tab:1}.

In Fig.~\ref{f5dune}, we display in dashed lines the expected 95\% CL sensitivity regions in the projection planes $|a_{\alpha \beta}|$ vs $|a_{\alpha \beta}^Z|$ and $(a_{\alpha \alpha}-a_{\tau \tau} )$ vs $(a_{\alpha \alpha}^Z-a_{\tau \tau}^Z)$, respectively, on the left and right panels. The projected 95\% CL sensitivities to the isotropic non-diagonal SME coefficients are $|a_{e \tau}|\lesssim 0.92\times 10^{-23}$ GeV $(|a_{e \tau}^Z|=0)$, $|a_{e \mu}|\lesssim 0.75\times 10^{-23}$ GeV $(|a_{e \mu}^Z|=0)$, and $|a_{\mu \tau}|\lesssim 0.68\times 10^{-23}$ GeV $(|a_{\mu \tau}^Z|=0)$, accordingly. However, for the spatial coefficients, the projected 95\% CL sensitivities are $|a_{e \tau}^Z|\lesssim 5.6 \times 10^{-23}$ GeV $(|a_{e \tau}|=0)$, $|a_{e \mu}^Z|\lesssim 4.7 \times 10^{-23}$ GeV $(|a_{e \mu}|=0)$, and $|a_{\mu \tau}^Z|\lesssim 4.5 \times 10^{-23}$ GeV $(|a_{\mu \tau}|=0)$, respectively.  

However, since the term $\hat{N}_Z$ is a fixed constant, for a single experimental configuration, we can redefine $a^T_{\alpha \beta}\rightarrow a^T_{\alpha \beta}-\hat{N}_Z a^Z_{\alpha \beta}$. For example, for the case of the DUNE-only setup, previous limits obtained in the literature on the isotropic SME coefficients $a^T_{\alpha \beta}$~\cite{Barenboim:2018ctx, Sarker:2023mlz, Raikwal:2023lzk, Agarwalla:2023wft} can now be interpreted as bounds on the combination: $a^T_{\alpha \beta}-\hat{N}_Z a^Z_{\alpha \beta}$.

In conclusion, in this Appendix we have explored the correlations among the isotropic and $Z-$spatial sectors at DUNE (Fig.~\ref{f5dune}). The limits on the isotropic $a_{\alpha \beta}$ can be relaxed depending on the limits set on the coefficients $a_{\alpha \beta}^{Z}$, for the $Z-$spatial coefficients. The projected 95\% CL sensitivities are $|a_{e \tau}^Z|\lesssim 5.6 \times 10^{-23}$ GeV, $|a_{e \mu}^Z|\lesssim 4.7 \times 10^{-23}$ GeV, and $|a_{\mu \tau}^Z|\lesssim 4.5 \times 10^{-23}$ GeV, accordingly (left panel of Fig.~\ref{f5dune}).

\end{document}